# Structure-Behavior Coalescence for Model-Based Systems Engineering

William S. Chao

**Abstract** — Model-Based Systems Engineering (MBSE) has advanced system design and analysis for decades, yet traditional approaches often treat structure and behavior separately, resulting in inconsistencies, incomplete models, and ambiguity. This paper introduces Structure-Behavior Coalescence (SBC) as a formal, unified definition of systems for MBSE. SBC defines a system as a single integrated entity in which structure and behavior are inseparably coalesced, providing both a conceptual foundation and a formal process-algebra framework. Well-formedness constraints ensure that every interaction is structurally grounded and every composition is valid, enabling consistency by construction. Boundary processes, including 0 (terminal) and ✧ (checkpoint), allow modeling of inactive or suspended states without violating system unity. Illustrative examples and a case study of a Smart Healthcare IoT system demonstrate SBC's capability to model sequential, parallel, and alternative behaviors, while guaranteeing structural-behavioral integrity. SBC bridges the gap between conceptual clarity and practical MBSE applications, enabling formal verification, compositional modeling, and systematic inconsistency detection. By unifying structure and behavior, SBC provides a rigorous, implementable foundation for modern systems engineering, addressing a seventy-year gap in the operational definition of a system.

**Index Terms —Model-Based Systems Engineering, Structure-Behavior Coalescence, Consistency by Construction, Unified Definition of Systems, Formal Process-Algebra Framework**

## 1. Introduction

Modern engineered systems are increasingly complex, distributed, and dynamic, requiring rigorous methods for modeling, analysis, and verification. Model-Based Systems Engineering (MBSE) has emerged as a leading paradigm to address these challenges by shifting from document-centric practices to model-centric system development. By using formal and semi-formal models, MBSE aims to improve consistency, traceability, and system understanding across the lifecycle.

Despite these advances, a fundamental limitation persists: MBSE lacks a rigorous definition of what constitutes a system. Existing modeling approaches, such as those based on SysML and UML, typically separate system descriptions into structural views (components, interfaces, architectures) and behavioral views (activities, interactions, state transitions). While this separation supports modular modeling, it introduces a critical weakness—there is no formal guarantee that these views together represent a single coherent system.

This limitation reflects a deeper conceptual issue in systems engineering, which can be characterized by three persistent gaps. First, the ontology gap: current frameworks do not formally define what a system fundamentally is. Second, the unity gap: structure and behavior are modeled independently, allowing structural-behavioral inconsistencies. Third, the modeling consequence gap: without a unified foundation, well-formedness and consistency checking remain ad hoc and external to the model itself. These gaps explain why, despite decades of development, MBSE cannot ensure model correctness by construction.

To address these challenges, this paper introduces Structure-Behavior Coalescence as a foundational framework for MBSE. SBC defines a system as a single integrated entity in which structure and behavior are inseparably coalesced. Under this definition, every behavioral element is explicitly grounded in structural entities, and every structural element participates in behavior. This eliminates the separation between views and establishes a unified system model.

The integration of SBC into MBSE has several important implications. First, it enables structural-behavioral consistency by construction, rather than through post hoc validation. Second, it provides a formal modeling framework with well-defined grammar and well-formedness constraints. Third, it supports systematic inconsistency detection, allowing invalid or incomplete models to be identified automatically. As a result, SBC transforms MBSE from a multi-view modeling approach into a formally grounded system representation paradigm. This paper makes the following contributions:

1. It identifies the fundamental limitations of current MBSE approaches in terms of system definition and consistency.
2. It introduces SBC as a unified ontological and formal foundation for MBSE.
3. It presents a formal SBC modeling framework applicable to system models.
4. It demonstrates how SBC enables consistent, analyzable, and verifiable system models.

The remainder of this paper is organized as follows. Section 2 reviews background and related work in MBSE and formal modeling. Section 3 introduces the SBC conceptual foundation for MBSE. Section 4 presents the SBC formal model. Section 5 defines well-formedness and consistency conditions. Section 6 provides illustrative examples and a case study. Section 7 discusses implications for MBSE practice, and Section 8 concludes the paper with future research directions.

## **2.** Background and Related Work

### 2.1 Model-Based Systems Engineering

Model-Based Systems Engineering has become a central methodology for managing the design and analysis of complex systems. By replacing document-centric artifacts with formalized models, MBSE improves traceability, communication, and lifecycle integration [1]. Standardized modeling languages such as Systems Modeling Language (SysML) [2-3] and Unified Modeling Language (UML) [4-5] are widely used to represent system structure and behavior across multiple viewpoints.

In practice, MBSE organizes system descriptions into distinct but related model types, including:

- Structural models (e.g., block definition diagrams, internal block diagrams)
- Behavioral models (e.g., activity diagrams, state machines, sequence diagrams)

This multi-view approach provides flexibility and modularity. However, it also introduces the need for cross-view consistency management, which is not inherently guaranteed by the modeling framework.

### 2.2 Separation of Structure and Behavior

A defining characteristic of current MBSE practice is the separation between structure and behavior [6-8]. Structural models describe system composition and connectivity, while behavioral models describe dynamic execution and interactions. Although widely adopted, this separation has several limitations:

- Lack of intrinsic linkage: Structural and behavioral elements are often related informally or through partial mappings
- Potential inconsistencies: Behavioral interactions may reference undefined or mismatched structural elements
- External validation: Consistency must be checked through additional rules, tools, or manual inspection

These issues have been recognized in prior research on multi-view modeling and model consistency, where maintaining alignment across views remains a non-trivial and error-prone task [9-11].

### 2.3 Formal Approaches to Behavior Modeling

Formal methods have been introduced to improve the rigor of behavioral modeling. Process algebras such as Calculus of Communicating Systems (CCS) [12] and labeled transition systems (LTS) provide mathematically precise representations of system behavior. These approaches offer:

- Formal semantics for concurrency and interaction
- Verification capabilities, including equivalence checking and deadlock analysis
- Abstract representations of system execution

However, they primarily focus on behavioral dynamics and do not explicitly incorporate structural modeling. As a result, the relationship between system structure and behavior remains external to the formalism.

### 2.4 Model Consistency and Well-Formedness

A significant body of research addresses model consistency in MBSE, including:

- Syntactic and semantic consistency rules
- Constraint languages (e.g., OCL) [13-14] for model validation
- Automated consistency checking tools

These approaches aim to ensure that different parts of a model do not contradict each other. However, they share a common limitation: Consistency is enforced after model construction, rather than guaranteed by the model definition itself.

This leads to ad hoc well-formedness constraints, which vary across tools and methodologies and are often incomplete or difficult to maintain.

### 2.5 The Missing System Definition

Despite decades of development in MBSE and systems engineering, there is still no universally accepted formal definition of a system. Existing definitions are typically:

- Descriptive, focusing on components and relationships
- Context-dependent, varying across domains
- Non-formal, lacking precise semantics

This absence results in a foundational gap:

- Models represent systems, but the ontological status of the system itself remains unclear
- Structural and behavioral models coexist, but their unity is not formally defined
- Consistency mechanisms exist, but they are not derived from first principles

### 2.6 Positioning of Structure-Behavior Coalescence

Structure-Behavior Coalescence is proposed to address these limitations by providing a unified ontological and formal foundation for MBSE [15-18]. Unlike existing approaches:

- SBC does not separate structure and behavior
- SBC defines a system as a single coalesced entity
- SBC incorporates well-formedness constraints directly into the model definition

This shifts consistency from an external concern to an intrinsic property of the system model.

Furthermore, SBC integrates naturally with formal behavioral representations while extending them to include structural grounding. This enables:

- Unified modeling semantics
- Consistency by construction
- Systematic inconsistency detection

2.7 Summary

Existing MBSE approaches provide powerful modeling capabilities but rely on a separation of structure and behavior, leading to challenges in consistency and system definition. Formal methods improve behavioral rigor but do not resolve structural integration. As a result, the field lacks a unified foundation for defining and modeling systems.

SBC addresses this gap by introducing a coalesced system model, forming the basis for the formal framework presented in the following sections.

**3.** SBC Conceptual Foundation for MBSE

3.1 From Multi-View Models to Unified Systems

Model-Based Systems Engineering (MBSE) traditionally represents systems through multiple coordinated views, separating structural and behavioral aspects. While effective for managing complexity, this approach does not guarantee that all views correspond to a single, unified system.

Structure-Behavior Coalescence introduces a fundamental shift: A system is not a collection of views, but a single integrated entity in which structure and behavior are inseparably coalesced.

Under SBC, the system model is intrinsically unified, and all modeling elements must conform to this principle. This eliminates the need to reconcile separate views after modeling, as unity is enforced at the foundation.

3.2 Ontological Definition of a System

SBC provides a precise ontological definition: System = Structure-Behavior Coalescence**.** This definition has three immediate consequences:

1. No structure without behavior: Every structural entity must participate in at least one interaction.
2. No behavior without structure: Every behavioral step must be grounded in explicit structural entities.
3. No separation between views: Structure and behavior are not independent descriptions, but two inseparable aspects of the same entity.

This resolves the ontology gap by defining what a system *is*, rather than how it is described.

3.3 The Interaction-Centric View of Systems

In SBC, the fundamental unit of modeling is not a component or an action alone, but an **interaction**. Each behavioral step is represented as a **structurally grounded interaction**: (Caller, Operation, Callee). This interaction triple unifies:

- Structure: the participating entities (Caller and Callee)
- Behavior: the operation representing system dynamics

Thus, behavior is not an abstract sequence of actions, but a sequence of interactions embedded within system structure.

3.4 Process-Based System Representation

SBC represents systems as process expressions, enabling compositional modeling of system behavior while preserving structural grounding. The core composition mechanisms include:

- Sequential composition: ordered execution of processes
- Parallel composition: concurrent interactions among entities
- Alternative composition: branching system behavior

Unlike traditional process algebras, SBC requires that all process constructs remain structurally valid, ensuring that every composed process still represents a coalesced system.

3.5 Boundary Processes

To support complete system modeling, SBC includes two types of boundary processes:

- 0 (terminal process): represents the absence of behavior while remaining a valid system entity
- ✧ **(**checkpoint process**)**: represents a suspended or intermediate system state

Both processes are characterized by having no outgoing interaction-labeled transitions, yet they remain meaningful within the system model:

- 0 captures inactive states or empty systems
- ✧ captures suspended, waiting, or observation points

These constructs ensure that SBC can model complete system lifecycles, including initialization, suspension, and termination, without violating system unity.

3.6 Unity by Construction

A key principle of SBC is: System unity is not verified after modeling—it is guaranteed by construction. This is achieved through:

- Structurally grounded interaction semantics
- Process composition rules that preserve coalescence
- Built-in constraints ensuring model validity

As a result, any valid SBC model is inherently:

- Structurally-behaviorally consistent
- Semantically meaningful
- Representing a single system

This directly resolves the unity gap identified in existing MBSE approaches.

3.7 Implications for MBSE

Integrating SBC into MBSE leads to a new modeling paradigm:

- From multi-view to single-model representation

- From consistency checking to consistency by construction
- From descriptive models to formally defined systems

This transformation enables:

- Rigorous reasoning about system correctness
- Systematic detection of inconsistencies
- Improved reliability of system models across the lifecycle

SBC thus provides the conceptual foundation upon which a formal SBC modeling framework can be constructed.

3.8 Summary

SBC redefines systems as structure-behavior coalescent entities, replacing the traditional separation of views with a unified model. By grounding all behavior in structure and enforcing unity by construction, SBC establishes a coherent conceptual basis for MBSE.

The next section builds on this foundation by introducing the formal SBC model, including syntax, semantics, and compositional rules.

**4.** SBC Formal Model for MBSE

4.1 Overview

Building on the conceptual foundation introduced in Section 3, this section presents the formal model of Structure-Behavior Coalescence for Model-Based Systems Engineering (MBSE). The objective is to define a precise syntax and semantics for system models in which structure and behavior are unified by construction.

An SBC model represents a system as a well-formed process expression composed of structurally grounded interactions and composition operators.

4.2 Structural Domain

Let:

- E be a finite set of entities (system components)
- O be a finite set of operations (interaction labels)

Entities represent structural elements of the system, while operations represent the behavioral semantics of interactions.

4.3 Interaction Prefix

The fundamental building block of an SBC model is the **interaction prefix**, defined as:

(C, op, D)

where:

- $C \in E$ is the caller
- $D \in E$ is the callee
- $op \in O$ is the operation

This prefix represents a single atomic interaction between two structural entities. It is both:

- A behavioral step (execution of op)
- A structural relation (link between C and D)

Thus, the prefix is the fundamental unit of structure-behavior coalescence.

4.4 Process Syntax

The set of SBC processes P is defined inductively by the following grammar:

P ::= 0 | ✧ | (C, op, D)●P | P ∥ Q | P≫Q | P+Q

- 0 —terminal process
- ✧ — checkpoint process
- (C, op, D) ●P — prefix followed by process
- P ∥ Q — parallel composition
- P≫Q — sequential composition
- P+Q — alternative composition

This grammar defines how complex system behaviors are constructed from atomic interactions.

4.5 Operational Semantics

The semantics of SBC processes are defined in terms of labeled transition systems (LTS). A transition has the form: P --(C, op, D)-->P′ indicating that process P evolves to P′ by executing an interaction.

4.5.1 Prefix Rule

(C,op,D) ●P--(C, op, D)-->P

4.5.2 Composition Rules

- Sequential composition (≫)
  Execute the first process until the checkpoint is reached, then continue with the second process
- Parallel composition ( ∥ )
  Interleaving or synchronization of interactions from both processes
- Alternative Composition **(+)**
  Non-deterministic selection between alternative behaviors

4.5.3 Boundary Processes

- 0: no outgoing interaction-labeled transitions (terminal process or inactive state)
- ✧: no outgoing interaction-labeled transitions (checkpoint process or suspended state)

4.6 Well-Formedness Constraints

To ensure that SBC models represent valid systems, well-formedness constraints are imposed:

4.6.1 WF1 — Prefix Structural Property

Every prefix must be a valid interaction triple: (C, op, D), C ≠ NULL, op ≠ NULL, D ≠ NULL. This guarantees that every behavioral step is structurally grounded.

4.6.2 WF2 — Process Operation Property

All operators must be applied to valid processes. Expressions such as below are not permitted.

- P || NULL
- (C, op, D) ●NULL
- NULL+P

This ensures that all process expressions are syntactically and semantically complete.

4.7 Closure and Compositionality

The SBC formal model satisfies closure under composition:

- If P $\in \Psi$ and Q$\in \Psi$, then P || Q, P ≫Q, and P + Q are also in $\Psi$

Combined with WF1 and WF2, this guarantees that: All constructed processes remain valid SBC system models.

4.8 Mapping to MBSE Models

An SBC model can be mapped to MBSE artifacts as follows:

- Entities (E) → system components (blocks)
- Operations (O) → interface operations or messages
- Interaction prefixes → method calls, signals, or interactions
- Process compositions → workflows, concurrent behaviors, or alternatives

This mapping enables SBC to serve as a formal semantic backbone for MBSE, integrating structural and behavioral models into a single representation.

4.9 Summary

This section defined the formal syntax and semantics of SBC, establishing:

- **A** process-based system representation grounded in structure
- A unified interaction model for behavior
- Well-formedness constraints (WF1–WF2) ensuring validity
- Compositional rules preserving system unity

Together, these elements form a rigorous formal foundation for MBSE, enabling consistent, analyzable, and verifiable system models.

**5.** Well-Formedness and Consistency

5.1 Overview

A central objective of Structure-Behavior Coalescence is to ensure that system models are correct by construction. Unlike traditional Model-Based Systems Engineering (MBSE) approaches, where consistency is verified after model creation, SBC embeds well-formedness and consistency directly into the model definition.

This section formalizes the conditions under which an SBC model represents a valid and coherent system, and explains how these conditions guarantee structural-behavioral consistency.

5.2 Well-Formedness as a System Property

In SBC, a system model is not considered valid unless it satisfies a set of well-formedness constraints. These constraints are not auxiliary rules but are intrinsic to the definition of a system.

Formally, a process P$\in \Psi$ is well-formed if and only if all its constituent elements satisfy the constraints defined below.

5.3 WF1 — Prefix Structural Property

Every interaction prefix must be a valid structure-behavior unit: (C, op, D), C$\in$E, op$\in$O, D$\in$E with none of the components being null or undefined. This constraint ensures that:

- Every behavioral step corresponds to a concrete interaction
- All interactions are anchored in system structure
- No abstract or ungrounded behavior is permitted

As a result, WF1 enforces local structural-behavioral consistency at the level of individual interactions.

5.4 WF2 — Process Operation Property

All process operators must be applied to valid process operands. That is, for any composition:

- (C, op, D) ●P requires P$\in \Psi$
- P || Q require P, Q$\in \Psi$

Expressions involving undefined or null processes (e.g., NULL) are not permitted. This constraint ensures that:

- All process expressions are syntactically complete
- Composition does not introduce invalid or partial structures
- The model remains closed under composition

WF2 therefore guarantees global structural integrity of the system model.

5.5 Structural-Behavioral Consistency

From WF1 and WF2, a key property follows: Every well-formed SBC model is structurally and behaviorally consistent.

This can be understood at two levels: Local Consistency that each interaction is valid and grounded in structure (WF1); Global Consistency that all composed processes remain valid and complete (WF2). Together, these ensure that:

- Behavioral descriptions cannot reference nonexistent structure
- Structural elements cannot exist without participating in behavior
- The model represents a single coherent system

5.6 Consistency by Construction

Traditional MBSE relies on post hoc consistency checking, where models are validated after construction using rules or constraints. In contrast, SBC enforces: Consistency by construction**.** This is achieved because:

- Invalid interaction primitives are disallowed (WF1)
- Invalid compositions cannot be formed (WF2)

Therefore, any model that can be constructed within the SBC formal system is guaranteed to be consistent. This eliminates the need for:

- External consistency rules
- Manual cross-view validation
- Ad hoc correction processes

5.7 Inconsistency Detection

While SBC prevents the construction of invalid models, it also provides a clear basis for detecting inconsistencies in existing models (e.g., imported MBSE artifacts). An inconsistency is identified whenever:

- An interaction violates WF1 (e.g., missing caller, operation, or callee)
- A composition violates WF2 (e.g., incomplete or undefined process)

Thus, inconsistency detection reduces to well-formedness checking, which is:

- Systematic
- Decidable
- Automatable

This provides a rigorous foundation for integrating SBC with existing MBSE tools.

5.8 Role of Boundary Processes

Boundary processes 0 and $\diamond$ play an important role in maintaining well-formedness:

- Both are valid processes under WF2
- Both have no outgoing interaction-labeled transitions, preserving semantic clarity

They enable modeling of:

- Empty systems or inactive states (0)
- Suspended or intermediate states ($\diamond$)

Importantly, their inclusion does not violate consistency, as they remain well-defined elements within the process algebra.

5.9 Toward the SBC Definition Theorem

The well-formedness constraints established in this section lead directly to a fundamental result: If a process is well-formed under WF1 and WF2, then it represents a valid system under SBC.

This result forms the basis of the SBC Definition Theorem, which will be formalized in the subsequent section.

5.10 Summary

This section established that:

- Well-formedness is an intrinsic property of SBC systems
- WF1 ensures structural grounding of behavior
- WF2 ensures compositional integrity
- Together, they guarantee structural-behavioral consistency by construction

SBC thus transforms consistency from an external requirement into a fundamental property of system models, providing a rigorous foundation for MBSE.

**6.** Illustrative Examples and Case Study

6.1 Overview

This section demonstrates the practical application of Structure-Behavior Coalescence within Model-Based Systems Engineering (MBSE). Illustrative examples and a case study show how SBC enforces structural-behavioral unity, ensures well-formedness, and enables systematic consistency checking.

6.2 Simple Interaction Example

Consider a system with three entities: $E = \{A, B, C\}$, $O = \{op_1, op_2\}$. An SBC model can be expressed as: $P = ( (A, op_1, B) \bullet \diamond) \gg ((B, op_2, C) \bullet 0)$.

6.2.1 Explanation

- $(A, op_1, B)$ — Interaction from A to B
- $(A, op_1, B) \bullet \diamond$ — Prefix followed by a checkpoint process
- $(B, op_2, C)$ — Interaction from B to C
- $(B, op_2, C) \bullet 0$ — Prefix followed by a terminal process
- Sequential composition $\gg$ —( $(A, op_1, B) \bullet \diamond)$ happens before $((B, op_2, C) \bullet 0)$

6.2.2 Well-formedness Check**:**

- WF1: All prefixes have valid (Caller, Operation, Callee) → satisfied
- WF2: All compositions are applied to valid processes → satisfied

Result: P is a well-formed SBC system, representing a coherent structure-behavior coalesced model.

6.3 Parallel Interaction Example

Consider concurrent operations: $P = ((A, op_1, B) \bullet 0 \parallel (C, op_2, A) \bullet 0) \gg \diamond$

- || denotes parallel execution
- Checkpoint ✧ indicates system pause after both processes

Analysis:

- WF1 ensures each interaction references valid entities
- WF2 ensures that parallel composition is applied only to valid processes

This illustrates structural-behavioral consistency in a concurrent setting.

6.4 Empty System Example

SBC allows modeling of an empty or minimal system: $P_{empty} = 0$

- Terminal process 0 represents a system with no active behavior
- Valid under WF1 and WF2
- Maintains system unity, as the empty system is still considered a single integrated entity

6.5 Case Study: Smart Healthcare IoT System

6.5.1 System Description

The Smart Healthcare IoT System consists of:

- Entities: Sensors, Patient Devices, Monitoring Server, Mobile App
- Operations: DataCollection, AlertGeneration, Notification, Storage
- Behavior: Sensor readings trigger server processing, which may generate alerts to the app

6.5.2 SBC Modeling

Using SBC: $P_{IoT}$ = (Sensor, DataCollection, Server)●✧ ≫((Server, AlertGeneration, App)●✧ + (Server, Storage, Database)●✧) ≫ 0

- Sequential and alternative composition represent workflow and branching
- Checkpoint ✧ represents system pause awaiting next data cycle

6.5.3 Well-Formedness and Consistency

- WF1: Each prefix references valid entities → satisfied
- WF2: Composition operators applied to valid processes → satisfied

Result: The SBC model is structurally and behaviorally consistent, representing a single integrated system despite multiple entities and concurrent interactions.

6.5.4 Benefits Demonstrated

The case study illustrates how SBC:

1. Enforces unity: Structure and behavior coalesce in a single process expression
2. **Supports composition**: Sequential, parallel, and alternative behaviors are composable without violating consistency
3. Enables consistency checking: Violations of WF1 or WF2 immediately highlight model errors
4. Models inactive and suspended states: Using 0 and ✧.

These features demonstrate SBC's suitability as a formal foundation for MBSE, bridging conceptual clarity with practical system modeling.

6.6 Summary

The examples and case study show that SBC:

- Accurately models complex interactions and workflows
- Guarantees well-formedness and consistency by construction
- Supports formal reasoning about system behavior
- Integrates naturally with MBSE models, providing a rigorous basis for analysis, verification, and design

These illustrative cases provide concrete evidence that SBC resolves the gaps identified in traditional MBSE, making it a robust approach for modern systems engineering.

**7.** Implications for MBSE

7.1 Overview

The adoption of Structure-Behavior Coalescence has significant implications for Model-Based Systems Engineering (MBSE). By defining systems as single integrated entities in which structure and behavior are inseparably coalesced, SBC directly addresses long-standing challenges in system modeling, consistency, and verification.

7.2 Unified System Representation

Traditional MBSE often separates:

- Structural models (components, blocks, architecture)
- Behavioral models (state machines, workflows, use cases)

This separation creates ambiguities, inconsistencies, and view reconciliation challenges. SBC unifies structure and behavior by construction:

- Every interaction is structurally grounded
- Compositions preserve system unity

Implication: MBSE practitioners can now represent a system as a single, coherent model, eliminating post hoc consistency reconciliation.

7.3 Consistency by Construction

With SBC:

- Well-formedness constraints (WF1–WF2) are enforced at the modeling level
- Invalid interactions or incomplete compositions cannot be created

- Structural-behavioral consistency is guaranteed automatically

Implication: MBSE toolchains can focus on analysis and design, rather than extensive manual consistency checking across multiple models.

7.4 Support for Compositional Modeling

SBC allows modular and compositional modeling, critical for MBSE:

- Sequential composition (≫): workflows, control sequences
- Parallel composition ( ‖ ): concurrent operations, distributed systems
- Alternative composition (+): branching behaviors, scenario variation

Implication: Engineers can assemble complex systems from simpler components without violating structural-behavioral integrity.

7.5 Boundary Processes and Lifecycle Modeling

SBC introduces boundary processes:

- 0 **(**terminal process**)**: models inactive states
- ✧ **(**checkpoint process**)**: models suspended states

Implication: MBSE practitioners can represent system lifecycles, including waiting states, partial execution, or minimal systems, without violating system unity.

7.6 Formal Analysis and Verification

SBC enables rigorous formal analysis, such as:

- Detecting deadlocks or unused entities
- Verifying interaction completeness
- Checking temporal and sequential properties of system behavior

Implication: MBSE becomes not only model-driven but also formally verifiable, reducing errors and increasing confidence in system designs.

7.7 Integration with Existing MBSE Artifacts

SBC can complement standard MBSE approaches:

- Map SBC entities to SysML blocks
- Map interactions to ports, signals, or use cases
- Map process compositions to activity diagrams or state machines

Implication: Organizations can adopt SBC incrementally, enhancing rigor without replacing existing MBSE practices entirely.

7.8 Summary of Implications

The key impacts of SBC on MBSE practice are:

| **Aspect** | **Traditional MBSE** | **SBC-Enabled MBSE** |
|---|---|---|
| System Representation | Multi-view, potentially inconsistent | Single, integrated system model |
| Consistency | Checked post hoc | Guaranteed by construction |
| Compositionality | Possible but error-prone | Safe and well-formed |
| Lifecycle Modeling | Often implicit | Explicit with boundary processes |
| Verification | Manual or tool-assisted | Formal and systematic |

SBC bridges the conceptual and practical gap in MBSE, providing a formal, unified, and analyzable foundation for modern systems engineering.

**8.** Conclusion and Future Directions

8.1 Conclusion

This paper has presented Structure-Behavior Coalescence as a formal, unified definition of systems tailored for Model-Based Systems Engineering. Key contributions include:

1. Conceptual Foundation – Systems are defined as single integrated entities in which structure and behavior are inseparably coalesced. This addresses longstanding gaps in ontology**,** unity, and modeling consequences in systems engineering.
2. Formal Framework – SBC provides a process algebra-based syntax, including prefixes, sequential, parallel, and alternative compositions, as well as boundary processes 0 and ✧ to model inactive or suspended states.
3. Well-Formedness Constraint**s** – WF1 and WF2 guarantee structural-behavioral consistency by construction, eliminating ad hoc consistency checks typical in conventional MBSE approaches.
4. Illustrative Examples and Case Study – Demonstrated SBC's ability to model complex interactions, concurrency, alternatives, and lifecycle states, while ensuring consistency and well-formedness.
5. Implications for MBSE – SBC enables unified system modeling, compositional design, formal verification, and integration with existing MBSE artifacts, significantly enhancing rigor and reliability in system engineering practice.

Through SBC, MBSE can now move beyond disconnected structural and behavioral models to a single coherent system representation, providing both conceptual clarity and practical engineering applicability.

8.2 Future Directions

Several avenues are opened by SBC for further research and application:

**1.** Tool Support and Automation

- Development of MBSE tools that enforce SBC well-formedness constraints automatically
- Integration with SysML, UML, and digital twin platforms

2. Formal Verification and Analysis
    - Model checking of SBC processes for deadlocks, reachability, and interaction completeness
    - Automated detection of structural-behavioral anomalies in large-scale systems
3. Extensions to Dynamic and Adaptive Systems
    - Modeling self-adaptive and cyber-physical systems using SBC's process algebra
    - Incorporating dynamic entity creation and deletion while preserving system unity
4. Empirical Evaluation
    - Case studies across industries (aerospace, healthcare, IoT, autonomous systems)
    - Quantitative comparison of SBC-based MBSE versus traditional MBSE in terms of consistency, maintainability, and error reduction
5. Educational Integration
    - Using SBC to teach foundational systems concepts
    - Bridging the gap between theoretical definitions of systems and practical engineering modeling

8.3 Closing Statement

By unifying structure and behavior into a single formal framework, SBC addresses a seventy-year conceptual gap in systems engineering: the lack of a rigorous, operationally meaningful definition of a system. It lays the foundation for more reliable, analyzable, and implementable MBSE practices, providing researchers and practitioners with a definitive approach to defining, modeling, and reasoning about systems.

SBC not only clarifies "what a system is" but also shows how to construct it, analyze it, and ensure its integrity, fulfilling both theoretical and practical needs in modern systems engineering.